\documentclass[amssymb,11pt]{article}
\usepackage[pdftex,colorlinks=true,urlcolor=black,filecolor=black,linkcolor=black,
            pdftitle={The Lanczos Potential for Bianchi Spacetime.},
            pdfauthor={Mark D. Roberts},
            pdfsubject={Cosmology},
            pdfkeywords={Lanczos Potential, Bianchi Spacetime},
            pagebackref,pdfpagemode=None,bookmarksopen=true]{hyperref}
\usepackage{amssymb}
\usepackage{graphicx}
\newcommand{\bc}{\begin{center}}
\newcommand{\ec}{\end{center}}
\newcommand{\be}{\begin{equation}}
\newcommand{\ee}{\end{equation}}
\newcommand{\ber}{\begin{eqnarray}}
\newcommand{\ear}{\end{eqnarray}}

\newcommand{\st}{\stackrel}

\begin{document}
\title{The Lanczos Potential for Bianchi Spacetime.}
\author{
\href{http://www.violinist.com/directory/bio.cfm?member=robemark}
{Mark D. Roberts},\\
}
\date{$27^{th}$ of September 2019}
\maketitle
\begin{abstract}
The form and application of the Lanczos potential for Bianchi spacetime are studied.
The Lanczos potential is found in some specific cases then the general case studied.
It leads to two coupled first order partial differential equations which although they
so far have not been solved in general can be solved for many configurations.
The application is to cosmic energetics:  in other words to the study of the energy of the
gravitational and other fields in the Universe.
\end{abstract}
{\small\tableofcontents}
\newpage
\section{Introduction.}\label{introduction}
The Lanzcos potential \cite{lanczos} discovered in (1962) is a potential tensor for the Weyl 
tensor in a similar manner to the vector potential $A$ being a potential for the 
Maxwell-Faradaty tensor $F$.   Just as for electromagnetism where one can form as stress 
energy tensor $T$ one can form an energy tensor (\ref{lpenergy}) \cite{mdr10}.
After finding exact expressions for the Lanczos potential and its form in general
whether associated scalars and energy expressions have application in the early or late 
universe is discussed.
The Lanzcos potential \cite{lanczos} is a first derivative tensor and so might have connection to 
properties of the covariant derivation of vector fields such as the shear and thus to chaotic and 
dynamical systems in csmology \cite{barrowjd,WE},  this is not looked at here.
The only application looked at here is to the energetics \cite{bel,mdr10} of the various spacetimes.

In \S\ref{lanczospotential} the Lanczos potential and its associated energy and analogs of the 
Weyl scalars are defined.
In \S\ref{schwarzschild} for illustration these objects are derived for Schwarzschild spacetime.
The Lanczos potential is known for Levi-Cevita spacetime and is reproduced in \S\ref{levicevita}.
In \S\ref{nonvaclevi} this is generalized for the case where the vacuum  constraints are not 
applied,  this is done by introducing a 'grand' shear tensor which is used in all subsequent
analysis.
In \S\ref{exposection} the properties of a spacetime with exponentials replacing the powers 
of Levi-Cevita is discussed.
The general case is discussed in \S\ref{bianchispacetime} where the general geometric objects
transvected Bel-Robinson tensor,  tranvected Lanczos potential energy and Lanczos scalars are given.
\S\ref{ce} applies the forgoing to speculation on the nature of gravitational energy in 
the Universe. 
\S\ref{conc} is the conclusion.
\section{The Lanczos Potential.}\label{lanczospotential}
The Weyl tensor can be expressed in terms of the Lanczos potential 
\begin{eqnarray}
\label{ldef}
\st{~}{C}_{abcd}&=&\st{1}{C}_{abcd}+\st{2}{C}_{abcd}+\st{3}{C}_{abcd}\\
\st{1}{C}_{abcd}&\equiv & H_{abc;d}-H_{abd;c}+H_{cda;b}-H_{cdb;a},~~
\st{3}{C}_{abcd}\equiv \frac{4}{(1-D)(2-D)}H^{ef}_{..e;f}(g_{ac}g_{bd}-g_{ad}g_{bc}),\nonumber\\
\st{2}{C}_{abcd}&\equiv & \frac{1}{(2-D)}\left\{g_{ac}(H_{bd}+H_{db})-g_{ad}(H_{bc}+H_{cb})+
    g_{bd}(H_{ac}+H_{ca})-g_{bc}(H_{ad}+H_{da})\right\},\nonumber
\end{eqnarray}
where the coefficients of $\st{2}{C}$ and $\st{3}{C}$ are fixed by requiring that the Weyl
tensor obeys the trace condition $C^{a}_{.bad}=0$.
The higher dimension equations were first given in \cite{mdrcs}.
$H_{bd}$ is defined by
\begin{equation}
\label{eq:6}
H_{bd}\equiv H^{~e}_{b.d;e}-H^{~e}_{b.e;d}.
\end{equation}
The Lanczos potential has the symmetries
\begin{equation}
\label{lpsym}
2H_{[ab]c}\equiv H_{abc}+H_{bac}=0,~~~
6H_{[abc]}\equiv H_{abc}+H_{bca}+H_{cab}=0.
\end{equation}
Equation (\ref{ldef}) is invariant under the algebraic gauge transformation
\begin{equation}
\label{alggaugetrans}
H_{abc}\rightarrow H'_{abc}=H_{abc}+\chi_a g_{bc}-\chi_b g_{ac},
\end{equation}
where $\chi_a$ is an arbitrary four vector,
this transformation again fixes the coefficients of $\st{2}{C}$ and $\st{3}{C}$ .

In four dimensions the Lanczos potential with the above symmetries has twenty degrees
of freedom,  but the Weyl tensor has ten.
Lanczos reduced the degrees of freedom to ten
by choosing the algebraic gauge condition
\begin{equation}
\label{alggaugecondition}
3\chi_a=H^{~b}_{a.b}=0,
\end{equation}
and the differential gauge condition
\begin{equation}
L_{ab}=H^{~~c}_{ab.;c}=0.
\label{ldg}
\end{equation}

Introducing a null tetrad $l,n,m,\bar{m}$ for signature $+++-$
\begin{eqnarray}
\label{nulltetrad}
&&1=-l\cdot n=-n\cdot l=m\cdot\bar{m}=\bar{m}\cdot m,\\
&&0=l^2=n^2=m^2=\bar{m}^2=l\cdot m=l\cdot\bar{m}=n\cdot m=n\cdot\bar{m},\nonumber\\
&&g_{ab}=-l_a n_b-n_a l_b+m_a\bar{m}_b+\bar{m}_am_b,\nonumber
\end{eqnarray}
the Weyl scalars are
\begin{eqnarray}
\label{weylscalars}
\Psi_0&\equiv&C_{a b c d}l{^a}m{^b}l{^c}m{^d},\\
\Psi_1&\equiv&C_{a b c d}l{^a}n{^b}l{^c}m{^d},\nonumber\\
\Psi_2&\equiv&C_{a b c d}l{^a}m{^b}\bar{m}{^c}n{^d},\nonumber\\
\Psi_3&\equiv&C_{a b c d}l{^a}n{^b}\bar{m}{^c}n{^d},\nonumber\\
\Psi_4&\equiv&C_{a b c d} n{^a}\bar{m}{^b}n{^c}\bar{m}{^d},\nonumber
\end{eqnarray}
there are similar objects for the Lamczos potential \cite{MZ,zund,odonnel}
\begin{eqnarray}
\label{hess}
H_0&\equiv& H_{abc}l^a m^b l^c,\\
H_1&\equiv& H_{abc}l^a m^b \bar{m}^c,\nonumber\\
H_2&\equiv& H_{abc}\bar{m}^a n^b l^c,\nonumber\\
H_3&\equiv& H_{abc}\bar{m}^a n^b \bar{m}^c,\nonumber\\
H_4&\equiv& H_{abc}l^a m^b m^c,\nonumber\\
H_5&\equiv& H_{abc}l^a m^b n^c,\nonumber\\
H_6&\equiv& H_{abc}\bar{m}^a n^b m^c,\nonumber\\
H_7&\equiv& H_{abc}\bar{m}^a n^b n^c,\nonumber
\end{eqnarray}

Energy tensors can be constructed from both the Weyl tensor and the Lanczos potential.
Define the dual to be on the first two indcies
\begin{equation}
\label{dualdef}
*H_{abc}\equiv\frac{1}{2}\epsilon_{abef}H^{ef}_{~..c},
\end{equation}
and similarly dualing over the first two indices for the Weyl tensor.
Then from the Weyl tensor one can construct the 
Bel-Robinson tensor \cite{bel}
\begin{equation}
\label{belrobinson}
B_{cdef}\equiv C_{acdb}C^{a~~b}_{.ef.}+*C_{acdb}*C^{a~~b}_{.ef.},
\end{equation}
which has dimensions energy squared,  given a until timelike vector field $V$ the energy
squared is 
\begin{equation}
\label{bvdef}
B_v\equiv B_{abcd}V^aV^bV^cV^d,
\end{equation}
which is of a similar form to the energy conditions \cite{HE} p.95;
in most cases looked at here it is proportional to the Weyl tensor squared 
${\rm WeylSq}=C_{abcd}C^{abcd}$.
Analogously from the Lanczos potential \cite{mdr10} ther eis the energy tensor
\begin{equation}
\label{lpenergy}
HE_{ab}\equiv H_{acd}H_{b..}^{~cd}+*H_{acd}*H_{b..}^{~cd},
\end{equation}
which has dimensions energy,  given a until timelike vector field $V$ the transvected energy is 
\begin{equation}
\label{hvdef}
H_v\equiv HE_{ab}V^aV^b,
\end{equation}
which has the correct dimensions of energy,  see also \cite{senovilla}.
It has the advantage that it gives a sign for the gravitational energy,
but has difficult interpretation as to whether it really is a measure of energy.
In many cases the transvected Bel-Robinson is proportional to the square to the Weyl
tensor $C_{abcd}C^{abcd}$ and in the Lanczos case proportional to $H_{abc}H^{abc}$.
\section{Schwarzschild Spacetime.}\label{schwarzschild}
In the most commonly used coordinates the Schwarzschild line element is
\begin{equation}
\label{schwatzschildle}
ds^2=\frac{dr^2}{1-\frac{2m}{r}}+r^2(d\theta^2+\sin(\theta)^2d\phi^2)
-\left(1-\frac{2m}{r}\right)dt^2.
\end{equation}
The Weyl tensor is given by
\begin{eqnarray}
\label{schweyl}
&C_{r\theta r\theta}=\frac{C_{r\phi r\phi}}{\sin(\theta)^2}=\frac{m}{2m-r},~~~
&C_{rtrt}=-\frac{2m}{r},\\
&C_{\theta t\theta t}=\frac{C_{\phi t \phi t}}{\sin(\theta)^2}=\frac{m(r-2m)}{r^2},~~~
&C_{\theta\phi\theta\phi\theta}=2mr\sin(\theta)^2,\nonumber
\end{eqnarray}
and its dual by
\begin{equation}
\label{schweyldual}
*C_{r\phi\theta t}=-*C_{r\theta\phi t}=\frac{1}{2}*C_{rt\theta\phi}=\frac{m}{r}\sin(\theta).
\end{equation}
The Lanczos potential is given by
\begin{eqnarray}
\label{schlp}
&H_{\theta r\theta}=\frac{H_{\phi r\phi}}{\sin(\theta)^2}=\frac{mr}{3(r-2m)},~~~
&H_{trr}=\frac{1}{r(r-2m)},\\
&H_{\theta t\theta}=\frac{H_{\phi t\phi}}{\sin(\theta)^2}=\frac{1}{2r},~~~
&H_{trt}=\frac{2m}{3r^2},
\end{eqnarray}
and its dual by
\begin{eqnarray}
\label{schlpdual}
&&*H_{\phi\theta r}=2*H_{\phi r\theta}=2*H_{r\theta\phi}=\frac{\sin(\theta)}{2m-r},\\
&&*H_{t\phi\theta}=*H_{\theta t\phi}=\frac{1}{2}*H_{\theta\phi t}=\frac{m}{3}\sin(\theta).
\nonumber
\end{eqnarray}
The energy squared and energy with respect to the unit timelike vector
\begin{equation}
\label{schwarzschildvec}
V_a=\left[0,0,0,\sqrt{1-\frac{2m}{r}}\right],
\end{equation}
are (\ref{bvdef},\ref{hvdef})
\begin{equation}
\label{bvsch}
B_v=\frac{6m^2}{r^6}=\frac{1}{8}{\rm WeylSq},~~~~~~
H_v=\frac{27-4m^2r^2}{18r(r-2m)},
\end{equation}
there is the relationship
\begin{equation}
\label{schodd}
H_1H_6=\frac{1}{24}H_{abc}H^{abc}=-\frac{1}{24}*H_{abc}*H^{abc}.
\end{equation}

Taking the tetrad
\begin{equation}
\label{schwarzschildtetrad}
l_a=\left[1,0,0,1-\frac{2m}{r}\right],~~~
n_a=\left[\frac{r}{2(2m-r)},0,0,\frac{1}{2}\right],~~~
m_a=\left[0,-\frac{{\mathit i}r}{\sqrt{2}},-\frac{r\sin(\theta)}{\sqrt{2}},0\right],
\end{equation}
the nonvanishing Weyl and Lanczos scalars are (\ref{weylscalars},\ref{hess})
\begin{equation}
\label{schscalars}
\Psi_2=-\frac{m}{r^3},~~~
H_1=\frac{3-2mr}{6r^3},~~~
H_6=\frac{3+2mr}{12r^2(2m-r)}.
\end{equation}
\section{Vacuum Levi-Cevita Spacetime.}\label{levicevita}
The line element is taken to be
\begin{equation}
\label{levicevitale}
ds^2=t^{2p_1}dx^2+t^{2p_2}dy^2+t^{2p_3}dz^2-dt^2.
\end{equation}
The Riemann,  tensor is
\begin{equation}
\label{levicevitarie}
R_{xyxy}=p_1p_2t^{2p_1+2p_2-2},~~~
R_{xtxt}=p_1(1-p_1)t^{2p_1-2},
\end{equation}
with $R_{xzxz},~R_{yzyz},~R_{ytyt},~R_{ztzt}$ following by symmetry.  The Ricci tensor is
\begin{equation}
\label{levicevitaric}
R_{xx}=p_1(p_1+p_2+p_3-1)t^{2p_1-2},~~
R_{tt}=\frac{1}{t^2}\left(p_1+p_2+p_3-p_1^2-p_2^2-p_3^2\right),\end{equation}
with $R_{yy},~R_{zz}$ following by symmetry.  The Ricci scalar is
\begin{equation}
\label{levicevitars}
R=\frac{2}{t^2}\left(-p_1-p_2-p_3+p_1p_2+p_1p_3+p_2p_3+p_1^2+p_2^2+p_3^2\right).
\end{equation}
The Weyl tensor is
\begin{eqnarray}
\label{levicevitaweyl}
&&C_{xyxy}=-\frac{1}{6}
\left(-p_1-p_2-p_3-2p_1p_2+p_1p_3+p_2p_3+p_1^2+p_2^2-2p_3^2\right)t^{2p_1+2p_2-2},\nonumber\\
&&C_{xtxt}=-\frac{1}{6}
\left(-2p_1+p_2+p_3-p_1p_2-p_1p_3+2p_2p_3+2p_1^2-p_2^2=p_3^2\right)t^{2p_1-2},
\end{eqnarray}
with $C_{xzxz},~C_{yzyz},~C_{ytyt},~C_{ztzt}$ following by symmetry.
The dual of the Weyl tensor is
\begin{equation}
\label{levicevitadw}
*C_{xyzt}=-\frac{1}{6}
\left(-p_1-p_2+2p_3-2p_1p_2+p_1p_3+p_2p_3+p_1^2+p_2^2-2p_3^2\right)t^{p_1+p_2+p_3-2},
\end{equation}
with $*C_{xzyt},~*C_{xtyz}$ following by symmetry.

From the Ricci tensor (\ref{levicevitaric}) for a vacuum
\begin{equation}
\label{vacequal}
p_1+p_2+p_3=p_1^2+p_2^2+p_3^2=1,
\end{equation}
the first equality in (\ref{vacequal}) gives the vacuum condition
\begin{equation}
\label{p3subs}
p_3=1-p_1-p_2,
\end{equation}
and the second equality in (\ref{vacequal}) gives the vacuum condition
\begin{equation}
\label{p2subs}
p_2=\frac{(1-p_1)}{2}\pm\frac{1}{2}{\rm rt_1},~~~
{\rm rt_1}\equiv\sqrt{(1-p_1)(1+3p_1^2)}.
\end{equation}
The transvected Bel-Robinson tensor (\ref{bvdef}) is
\begin{eqnarray}
\label{levicevitabv}
B_v&=&\frac{1}{8}{\rm WeylSq}\\
&=&\frac{1}{6t^4}\left[p_1^4+p_2^4+p_3^4-p_1^3(p_2+p_3+2)-p_2^3(p_1+p_3+2)-p_3^3(p_1+p_2+2)\right.\nonumber\\
&&\left.+p_1^2(p_2p_3+2(p_2+p_3)+1)+p_2^2(p_1p_3+2(p_1+p_3)+1)+p_3^2(p_1p_2+2(p_1+p_2)+1)\right.\nonumber\\
&&\left.-6p_1p_2p_3-p_1p_2-p_1p_3-p_2p_3\right],\nonumber
\end{eqnarray}
imposing (\ref{p3subs},\ref{p2subs}),  (\ref{levicevitabv}) reduces to
\begin{equation}
\label{levvacbv}
B_v=\frac{2(1-p_1)p_1^2}{t^4}.
\end{equation}

The unit timelike vector field is
\begin{equation}
\label{levicevitavf}
V_a=\left[0,0,0,1\right],
\end{equation}
which without imposing (\ref{p3subs},\ref{p2subs}) it has shear
\begin{equation}
\label{levicevitashear}
\sigma_{ab}=\frac{1}{3}{\rm diag}
\left\{(-2p_1+p_2+p_3)t^{2p_1-1},(p_1-2p_2+p_3)t^{2p_2-1},(p_1+p_2-2p_3)t^{2p_3-1},0\right\}.
\end{equation}
The Lanczos potential,  compare \cite{NV},  is
\begin{equation}
\label{levicevitalp}
H_{abc}=\frac{1}{3}\left(\sigma_{a c}V_b-\sigma_{bc}V_a\right),
\end{equation}
provided the vacuum condition (\ref{p3subs}) is imposed,  
it is not necessary to impose (\ref{p2subs}).
The energy (\ref{hvdef}) is
\begin{eqnarray}
\label{lcen}
H_v&=&\frac{2}{27t^2}\left(p_1^2+p_2^2+p_3^2-p_1p_2-p_1p_3-p_2p_3\right)\\
  &=&\frac{4}{27}S^t_{.t}+\frac{2}{27t^2}(p_1+p_2+p_3),\nonumber
\end{eqnarray}
where $S_{ab}$ is the traceless Ricci tensor,
imposing (\ref{p3subs},\ref{p2subs}) this becomes
\begin{equation}
\label{lcvacen}
H_v=\frac{2}{27t^2},
\end{equation}
which is positive and independent of $p$.

Choosing the tetrad compare \cite{CBBP}
\begin{equation}
\label{vactet}
l_a=[\sqrt{g_{xx}},0,0,1]/\sqrt{2},~~~
n_a=[-\sqrt{g_{xx}},0,0,1]/\sqrt{2},~~~
m_a=[0,\sqrt{g_{yy}},{\mathit i}\sqrt{g_{zz}},0]/\sqrt{2},
\end{equation}
and imposing (\ref{p3subs},\ref{p2subs}),  the Weyl scalars (\ref{weylscalars}) are
\begin{equation}
\label{vacweyls}
\Psi_0=4\Psi_4=-\frac{p_1}{2t^2}{\rm rt_1},~~~
\Psi_2=\frac{p_1(1-p_1)}{t^2},
\end{equation}
the Lanczos scalars (\ref{hess} are
\begin{equation}
\label{vaclac}
H_1=-H_6=\frac{(3p_1-1)}{18\sqrt{2}t},~~~
H_4=-2H_3=\frac{{\rm rt_1}}{6\sqrt{2}t},
\end{equation}
with $rt_1$ defined in (\ref{p2subs}).
\section{Non-vacuum Levi-Cevita.}\label{nonvaclevi}
For the line element (\ref{levicevitale}) without imposing the second constraint (\ref{p2subs})
it is necessary to add a new diagonal tensor,  here called the 'grand' shear, to the shear (\ref{levicevitashear})
\begin{equation}
\label{newlc}
\Sigma_{ab}={\rm diag}
\left\{f_1(t)g_{xx},f_2(t)g_{yy},f_3(t)g_{zz},0\right\},
\end{equation}
where the explicit dependence on $t$ is left out when the ellipsis is clear,  i.e. $f_1(t)\rightarrow f_1$.
the algebraic gauge condition (\ref{alggaugecondition}) entails that the tensor is tracless
here realized by $f_1=-f_2-f_3$.
For the line element (\ref{levicevitale}) the $f$'s are given by
\begin{eqnarray}
\label{nvlcfs}
&&f_2=\frac{(p_1+p_2+p_3-1)}{6(d-1)t}\left(p_1^2-2p_2^2+p_3^2+2p_1p_2-4p_1p_3+2p_2p_3+p_1-2p_2+p_3\right),\nonumber\\
&&f_3=\frac{(p_1+p_2+p_3-1)}{6(d-1)t}\left(p_1^2+p_2^2-2p_3^2-4p_1p_2+2p_1p_3+2p_2p_3+p_1+p_2-2p_3\right),\nonumber\\
&&d\equiv(p_1^2+p_2^2+p_3^2-p_1p_2-p_1p_3-p_2p_3),
\end{eqnarray}
to which can be added the non-contributing terms
\begin{eqnarray}
\label{nvlsnc}
&&f_{2C}=C_1\cos\left(\sqrt{-d-2}\ln(t)\right)+C_2\sin\left(\sqrt{-d-2}\ln(t)\right),\nonumber\\
&&f_{3C}=C_3\cos\left(\sqrt{-d-2}\ln(t)\right)+C_4\sin\left(\sqrt{-d-2}\ln(t)\right),\nonumber\\
&&C_3=\frac{1}{p_1-p_3}\left((p_3-p_2)C_1-\sqrt{d-2}C_2\right),~~~
C_4=\frac{1}{p_1-p_3}\left((p_3-p_2)C_2+\sqrt{d-2}C_1\right),\nonumber
\end{eqnarray}
where $C_1,~C_2,~C_3,~C_4$ are constants,  these terms do not explicitly contribute to the Weyl 
tensor via (\ref{ldef}) but cancel out.
The grand shear's {\ref{newlc}) properties are that it it is tracefree,  this is a consquence 
of imposing the algebraic gauge condition (\ref{alggaugecondition}),  and that taking the Lie 
derivative with respect to the unit timelike vector field (\ref{levicevitavf}) gives
\begin{equation}
\label{tdnewlc}
{\mathcal L}_v\Sigma_{ab}\equiv\Sigma_{ab;c}V^c=\frac{1}{t}\Sigma_{ab}.
\end{equation}
Another property is that (\ref{newlc}) can be reduced to a covariant derivative of a vector 
field which in turn can be thought of as a potential compare \cite{udeschihi},  
but as this is not useful for present purposes so we stay with (\ref{newlc}).
The Bel-Robinson tensor is still given by (\ref{levicevitabv}).
Now the Lanczos potential
(\ref{levicevitalp}) has an additional term
\begin{equation}
\label{levicevitanonv}
H_{abc}=\frac{1}{3}\left((k\sigma_{a c}+\Sigma_{a c})V_b-(k\sigma_{bc}+\Sigma_{bc})V_a\right),
\end{equation}
where except for \S\ref{bianchispacetime} $k=1$.
The Lanczos potential transvected energy (\ref{hvdef}) is
\begin{eqnarray}
\label{lcvacen2}
H_v&=&
\frac{1}{6(d-1)^2t^2}\left[+(p_1^6+p_2^6+p_3^6)-2((p_2+p_3)p_1^5+(p_1+p_3)p_2^5+(p_1+p_2)p_3^5)\right.\nonumber\\
&&+(2p_2^2+2p_3^2+2p_2p_3+p_2+p_3-2)p_1^4
  +(2p_1^2+2p_3^2+2p_1p_3+p_2+p_3-2)p_2^4\nonumber\\
&&+(2p_1^2+2p_2^2+2p_1p_3+p_1+p_2-2)p_3^4-2(p_1^3p_2^3+p_1^3+p_2^3p_3^3)\nonumber\\
&&+(-p_2^2-p_3^2-4p_2p_3+3p_2+3p_3)p_1^3
  +(-p_1^2-p_3^2-4p_1p_3+3p_1+3p_3)p_2^3\nonumber\\
&&+(-p_1^2-p_2^2-4p_1p_2+3p_1+3p_2)p_3^3
-3p_1^2p_2^2p_3^2
+4(p_2p_1^2p_3^2+p_3p_1^2p_2^2+p_1p_2^2p_3^2)\nonumber\\
&&+6p_1p_2p_3
-(p1^2p_3^3+p_2^2p_3^2+p_1^2p_3^2++3p_2p_3p_1^2+3p_1p_3p_2^2+p_1p_2p_3^2)\nonumber\\
&&+(1-p_2-p_3)p_1^2+(1-p_1-p_3)p_2^2+(1+p_1+p_2)p_3^2\nonumber\\
&&\left.-p_1p_2-p_1p_3-p_2p_3\right],
\end{eqnarray}
with $d$ defined in (\ref{nvlcfs}).

Using the same null tetrad (\ref{vactet}) as before the Weyl and Ricci scalars are
\begin{eqnarray}
\label{nvacweylriciis}
&&\Psi_0=4\Psi_4=\Phi_{02}=\Phi_{20}=\frac{(p_2-p_3)(p_1-p_2-p_3+1)}{4t^2},\\
&&\Psi_2=\frac{1}{12t^2}\left(-2p_1^2+p_2^2+p_3^2+p_1p_3+p_1p_3-2p_2p_3+2p_1-p_2-p_3\right),\nonumber\\
&&\Phi_{00}=\Phi_{22}=\frac{(p_1p_2+p_2p_3+p_2-p_2^2+p_3-p_3^2)}{4t^2},~~~
\Phi_{11}=\frac{(-p_1^2+p_1+p_2p_3)}{4t^2},\nonumber
\end{eqnarray}
the Lanczos scalars are
\begin{eqnarray}
\label{nvlcls}
H_1&=&-H_6=-\frac{1}{2\sqrt{2}(d-1)}\times\\
&&[-2p_1^3+p_2^3+p_3^2+2(p_2+p_3)p_1^2-(p_1+p_3)p_2^2-(p_1+p_2)p_3^2\nonumber\\
&&~~~-p_1p_2-p_1p_3+2p_2p_3-2p_1-p_2-p_3],\nonumber\\
H_4&=&-H_3=\frac{3}{2\sqrt{2}(d-1}\left((p_2-p_3)(p_2^2+p_3^2-(p_2+p_3+1)p_1-1\right),\nonumber
\end{eqnarray}
with $d$ defined in (\ref{nvlcfs}).

In order to investigate how the vacuum Levi-Cevita 'sits' in the non-vacuum transform the $p$ 
constants to $\epsilon$ constants
\begin{eqnarray}
\label{changevar}
&&\epsilon_2\equiv p_1+p_2+p_3-1,~~~
\epsilon_3\equiv p_1^2+p_2^2+p_3^2-1,\\
&&p_2=\frac{1}{2}(1+\epsilon_2-p_1)+\frac{1}{2}rt_2,~~~
p_3=\frac{1}{2}(1+\epsilon_2-p_1)-\frac{1}{2}rt_2,\nonumber\\
&&rt_2^2\equiv(1-p_1)(3p_1+1)+\epsilon_2(-\epsilon_2-2(1-p_1))+2\epsilon_3,\nonumber
\end{eqnarray}
and because there is only one $p$ left drop the index on $p_1\rightarrow p$.
Despite the $\epsilon$ notation all the following are exact expressions,
no expansions are used.
Firstly look at the $\epsilon_2=0$ case,  calculations are most easily achieved applying this
at the last moment as the are expressions with $\epsilon_2$ in both numerator and denominator,
the equations (\ref{levicevitabv},\ref{lcvacen},\ref{nvacweylriciis},\ref{nvlcls}) reduce to
\begin{eqnarray}
\label{e20}
&&B_v=\frac{1}{8}WeylSq=\frac{1}{6t^4}\left(12p(1-p)+6p\epsilon_3+\epsilon_3^2\right),~~~
H_v=\frac{2+3\epsilon_3}{27t^2},\nonumber\\
&&\Psi_0=\Psi_4=\frac{6p}{\sqrt{2}t}H_3=-\frac{6p}{\sqrt{2}t}H_3
=\frac{p}{2t^2}\sqrt{1+2p-3p^2+2\epsilon_3},\nonumber\\
&&\Psi_2=-\frac{1}{12t^2}\left(6p(p-1)-2\epsilon_3\right),~~~
\Phi_{00}=\Phi_{22}=2\Phi_{11}=6\Phi_l=-\frac{\epsilon_3}{4t^2},\nonumber\\
&&H_1=-H_6=\frac{\sqrt{2}}{36t}(3p-1),
\end{eqnarray}
$H_v$ remains explicitly independent of $p$ as in the vacuum case (\ref{lcvacen}),  
althought there can be thought of as an implicit dependence via $\epsilon_3$,
for $\epsilon_3<-2/3$ it changes sign.
Secondly look at the $\epsilon_3=0$ case,  the equations 
(\ref{levicevitabv},\ref{lcvacen},\ref{nvacweylriciis},\ref{nvlcls}) reduce to
\begin{eqnarray}
\label{e30}
&&B_v=\frac{2(1-p)p^2}{t^4}+\frac{p(p+2)(p-1)}{t^4}\epsilon_2
+\frac{1-2p^2}{2t^4}\epsilon_2^2+\frac{p}{2t^4}\epsilon_2^3-\frac{1}{12t^2}\epsilon_2^4,\nonumber\\
&&H_v=\frac{1}{6(2+\epsilon_2)^2t^2}
\left[12p^2(1-p)-12p(p-1)^2\epsilon_2+(4-18p+12p^2)\epsilon_2^2+2(2-3p)\epsilon_2^3+\epsilon_2^4\right],\nonumber\\
&&\Psi_2=-\frac{1}{12t^2}\left[6p(p-1)+3(1-p)\epsilon_2+\epsilon_2^2\right],~~~
\Psi_0=\Psi_4=\frac{2p-\epsilon_2}{4t^2}rt_3,\nonumber\\
&&\Phi_{00}=\Phi_{22}=\frac{1+p}{4t^2}\epsilon_2,~~~
\Phi_{11}=\frac{(2(1-p)+\epsilon_2)}{8t^2}\epsilon_2,~~~
\Phi_l=\frac{1}{24t^2}\epsilon_2^2,\nonumber\\
&&H_1=-H_6=-\frac{\sqrt{2}}{12(2+\epsilon_2)t}\left[3p(p-1)+(2-3p)\epsilon_2+\epsilon_2^2\right],~~~
H_3=-H_4=\frac{\sqrt{2}p}{4(2+\epsilon_2)t}rt_3,\nonumber\\
&&rt_3^2\equiv1+2p-3p^2+2(p-1)\epsilon_2-\epsilon_2^2
\end{eqnarray}
and in this case (\ref{lcvacen}) is not explicitly independent of $p$.
The significance of (\ref{e20},\ref{e30}) is discussed in the conclusion \S\ref{conc}.
\section{Exponential Spacetime.}\label{exposection}
The line element is taken to be
\begin{equation}
\label{expyspacetime}
ds^2=\exp(2p_1t)dx^2+\exp(2p_2t)dy^2+\exp(2p_3t)dz^2-dt^2.
\end{equation}
The Riemann and Ricci tensors are
\begin{eqnarray}
\label{expyrie}
&&R_{xyxy}=p_1p_2\exp(2(p_1+p_2)t),~~~~~~~
R_{xtxt}=-p_1^2\exp(2p_1t),\nonumber\\
&&R_{xx}=p_1(p_1+p_2+p_3)\exp(2p_1t),~~~~
R_{tt}=-p_1^2-p_2^2-p_3^2,\nonumber\\
&&R=2(p_1^2+p_2^2+p_3^2+p_1p_2+p_1p_3+p_2p_3),
\end{eqnarray}
with $R_{xzxz},~R_{yzyz},~ R_{ytyt},~R_{ztzt},R_{yy},~R_{zz}$ following by symmetry.
the combination of $p$'s for a vacuum is different than for Levi-Cevita also the tensors
can be expressed indepently of $t$.  
The $p$'s can be though of as three 'cosmological constants'.
The Weyl tensor is
\begin{eqnarray}
\label{expyweyl}
&&C_{xyxy}=-\frac{1}{6}\exp(2(p_1+p_2)t)\left(p_1^2+p_2^2-2p_3^2-2p_1p_2+p_1p_3+p_2p_3\right),\nonumber\\
&&C_{xtxt}=-\frac{1}{6}\exp(2p_1t)\left(2p_1^2-p_2^2-p_3^2-p_1p_2-p_1p_3+2p_2p_3\right),\\
&&*C_{xyzt}=-\frac{1}{6}\exp((p_1+p_2+p_3)t)\left(p_1^2+p_2^2-2p_3^2-2p_1p_2+p_1p_3+p_2p_3\right),\nonumber
\end{eqnarray}
with $C_{xzxz},~C_{yzyz},~ C_{ytyt},~C_{ztzt},~*C_{xzyt},~*C_{xtyz}$ following by symmetry.

For the line element (\ref{expyspacetime}) the $f$'s are given by
\begin{eqnarray}
\label{expy}
f_2=\frac{p_1+p_2+p_3}{6(d-1)}\left(p_1^2-2p_2^2+p_3^2+2p_1p_2-4p_1p_3+2p_2p_3\right),\nonumber\\
f_3=\frac{p_1+p_2+p_3}{6(d-1)}\left(p_1^2+p_2^2-2p_3^2-4p_1p_2+2p_1p_3+2p_2p_3\right),
\end{eqnarray}
to which can be added the non-contributory terms
\begin{eqnarray}
\label{expyaux}
&&f_{2C}\equiv C_1\cos\left(\sqrt{1-d}t\right)+C_2\sin\left(\sqrt{1-d}\right),\\
&&f_{3C}\equiv C_3\cos\left(\sqrt{(1-d)}t\right)+C_4\sin\left(\sqrt{(1-d)}\right),\nonumber\\
&&C_1\equiv\frac{1}{p_2-p_1}\left((p_3p_2)C_3+\sqrt{1-d}C_4\right),~~
C_2\equiv\frac{1}{p_2-p_1}\left((p_3p_2)C_4-\sqrt{1-d}C_3\right),\nonumber
\end{eqnarray}
The Lie derivative of this with respect to the unit timelike vector
field (\ref{levicevitavf}) vanishes,  compare (\ref{tdnewlc}).
The expression (\ref{levicevitanonv}) 
gives the Lanczos potential.
The transvected Bel-Robinson (\ref{bvdef}) and transvected Lanczos energy (\ref{hvdef}) are
\begin{equation}
\label{expyv}
B_v=d_2H_v=\frac{1}{6}
\left(p_1^4+p_2^4+p_3^4-p_1^3(p_2+p_3)-p_2^3(p_1+p_3)-p_3^3(p_1+p_2)+p_1p_2p_3(p_1+p_2+p_3)\right)
\end{equation}

The Weyl,  Ricci  and Lanczos scalars are
\begin{eqnarray}
\label{expywr}
&&\Psi_0=\Psi_4=\Phi_{02}=\Phi_{20}=\frac{1}{4}(p_2-p_3)(p_1-p_2-p_3),\nonumber\\
&&\Psi_2=\frac{1}{12}\left(-2p_1^2+p_2^2+p_3^2+p_1p_2+p_1p_3-2p_2p_3\right)\nonumber\\
&&\Phi_{00}=\Phi_{22}=\frac{1}{4}(p_1p_2+p-2p_3-p_2^2-p_3^2),~~~~~
\Phi_{11}=\frac{1}{4}(-p_1^2+p_2p_3),\nonumber\\
&&H_1=-H_6=-\frac{\sqrt{2}}{24(d-1)}\left(-2p_1^3+p_2^3+p_3^3+2p_1(p_2+p_3)-p_2^2(p_1+p_3)-p_3(p_1+p_2)\right),\nonumber\\
&&H_3=-H_4=-\frac{\sqrt{2}}{8(d-1)}(p_3-p_2)\left(p_2^2+p_3^2-p_1(p_2+p_3)\right).
\end{eqnarray}
\section{The general case.}\label{bianchispacetime}
Take line element of the form
\begin{equation}
\label{BI}
ds^2=A_1(t)^2dx^2+A_2(t)^2dy^2+A_3(t)dz^2-dt^2,
\end{equation}
where the explicit dependence on $t$ is left out when the ellipsis is clear,  
i.e. $A_1(t)\rightarrow A_1$.
As before take unit timelike vector field (\ref{levicevitavf}),  grand shear (\ref{newlc}),
and Lanczos potential (\ref{levicevitanonv}) with $k$ now not necessarily $k=1$.
The Lie derivative of the grand shear (\ref{newlc}) along (\ref{levicevitavf}) is
\begin{equation}
\label{liegen}
{\mathcal L}_v\Sigma_{ab}={\rm diag}
\left\{-\dot{f_1}(t)g_{xx},-\dot{f_2}(t)g_{yy},-\dot{f_3}(t)g_{zz},0\right\},
\end{equation}
so it is only in restricted cases that equations like (\ref{tdnewlc}) happen.
Define
\begin{equation}
\label{defbeta}
\beta_I\equiv\frac{\dot{A}_I}{A_I},
\end{equation}
where the index $I$ is not summed.
The effect of $k\ne1$ is the same as the transformations
\begin{equation}
\label{transformation}
f_2\rightarrow f_2+\frac{k}{3}\left(\beta_1-2\beta_2+\beta_3\right),~~~
f_3\rightarrow f_3+\frac{k}{3}\left(\beta_1+\beta_2-2\beta_3\right).
\end{equation} 
For (\ref{levicevitanonv}) to be a Lanczos potential substituting in (\ref{ldef})
gives the coupled partial differential equations
\begin{eqnarray}
\label{grandequ}
&&\gamma_2+(-\beta_2+\beta_3)f_2+(-\beta_1+\beta_3)f_3-\dot{f_2}=0,\nonumber\\
&&\gamma_3+(-\beta_1+\beta_2)f_2+(\beta_2-\beta_3)f_3-\dot{f_3}=0,
\end{eqnarray}
where
\begin{eqnarray}
\label{gammadef}
6\gamma_2\equiv&&(3-2k)\left(\dot{\beta_1}+\beta_1^2-2\dot{\beta_2}-2\beta_2^2+\dot{\beta_3}+\beta_3^2\right)\nonumber\\
&&+(3-4k)\left(\beta_1\beta_2-2\beta_1\beta_3+\beta_2\beta_3\right),\nonumber\\
6\gamma_3\equiv&&(3-2k)\left(\dot{\beta_1}+\beta_1^2+\dot{\beta_2}+\beta_2^2-2\dot{\beta_3}-2\beta_3^2\right)\nonumber\\
&&+(3-4k)\left(-2\beta_1\beta_2+\beta_1\beta_3+\beta_2\beta_3\right),
\end{eqnarray}
as $k$ is free it can be set to remove either the first $k=3/2$ or second $k=3/4$ terms 
in $\delta$.
In the general case the equations remain intractable,  exceptions being $f$'s a constant 
which is the case of \S\ref{exposection} and the $f$'s are proportional to $1/t$ which is the 
case of \S\ref{nonvaclevi}.
The equations (\ref{grandequ}) seem to be the most general because altering them gives functions
which are absorable by the $f$'s..

Seven possible approaches:
\begin{enumerate}
\item
\label{go}
go for the general case $f(\beta)$ but this has proved intractable so far,
\item
\label{trac}
consider if the equations are tractable for some examples in the Bianchi 
classification,  which is left for now,
\item
\label{compat}
see if known systems of pdes are compatible with (\ref{grandequ}),
\item
\label{choosea}
choose $A$ find $f$,  see the next paragraph,
\item
\label{choosef}
choose $f$ as either a function of $\beta$ and/or $t$ then find $A$, 
see the paragrapgh after next,
\item
\label{newfieldeq}
choose new field equations such as
\begin{equation}
\label{newfldeq}
HE_{a b}=\kappa G_{a b},
\end{equation}
where $HE_{ab}$ is given by (\ref{lpenergy}) with interpretation that the matter and 
gravitational energy are proportional,  there are other possibilities such as G proporional 
to the grand shear $\Sigma$ (\ref{levicevitanonv}),
\item
\label{qual}
ignore,  rather than exact solutions for $f$,  qualitative properties such as 
zeros and sign are what are important.
\end{enumerate}

To illustrate \ref{choosea}) for
\begin{equation}
\label{trigexample}
A_1(t)=a_1\cos(ct),~~~
A_2(t)=a_2\sin(ct),~~~
A_3(t)=a_3\sin(ct),
\end{equation}
solving (\ref{grandequ}) gives
\begin{eqnarray}
\label{trigsol}
&&f_2=-\frac{c}{4\cos(ct)\sin(ct)}+C_3\tan(ct)+{\mathit i}C_1\sinh(\mu)+C_2\cosh(\mu),\\
&&f_3=-\frac{c}{4\cos(ct)\sin(ct)}+C_3\tan(ct)-C_2\sinh(\mu)-{\mathit i}C_1\cosh(\mu),\nonumber
\end{eqnarray}
where
\begin{equation}
\label{defmu}
\mu(t)\equiv\ln\left(\frac{1-\cos(ct)-\sin(ct)}{\sin(ct)}\right)
+\ln\left(\frac{1-\cos(ct)+\sin(ct)}{\sin(ct)}\right)
-\ln\left(\frac{\sin(ct)}{1+\cos(ct)}\right),
\end{equation}
taking $k=C_1=C_2=C_3=0$ leaves just the first terms in (\ref{trigsol}) which is sufficient to 
cover the Weyl tensor;  $C_1=C_2=C_3\ne0$ might have application in quantum theory.

To illustrate \ref{choosef}),  
Levi-Cevita \S\ref{levicevita} can be thought of as $1/t$ Bianchi spacetime
and exponential spacetime \S\ref{exposection} as constant,  so what happens for $t^n$,
for example take $n=1$ 
\begin{equation}
\label{fpowert}
f_2(t)=k_2t,~~~f_3(t)=k_3t,
\end{equation}
(\ref{grandequ}) has solution
\begin{eqnarray}
\label{fpowersol}
&&\beta_1=-\frac{2}{3}(2k_2+k_3)t,~~~
\beta_3=-\frac{1}{3t}(3+2(k_3+2k_2)t^2),\\
&&\beta_2=-\frac{2t(k_3+2k_2)(-3+2(k_2+2k_3)t^2}{9+6(k_2+2k_3)t^2}\nonumber
\end{eqnarray}
integrating and exponentiating
\begin{equation}
\label{fpoweras}
A_1=\exp\left\{-\frac{1}{3}(2k_2+k_3)t^2\right\},~~~
A_2=\alpha_2(t)A1,~~~
A_3=\frac{A_1}{t},
\end{equation}
where
\begin{eqnarray}
\label{defalpha2}
&&\alpha_2=\left(3+2(k_2+2k_3)t^2\right)^\frac{2k_2+k_2}{k_2+2k_3},~~~~~~~~~~~~~~~~~~~~~~~~{\rm for}~~~k=\frac{3}{2},\\
&&\alpha_2=t^\frac{k}{2k-3}
\exp\left(\frac{(k_2+2k_3)t^2}{2(2k-3)}\right)\times \nonumber\\
&&~~~~~~\left\{C_1WM(w_1,w_2,w_3)+C_2WW(w_1,w_2,w_3)\right\}
~~~{\rm for}~~~k\ne\frac{3}{2},\nonumber\\
&&w_1\equiv\frac{(7k/2-6)k_2+(k-3)k_3}{(2k-3)(k_2+2k_3)},~~~
w_2\equiv\frac{\sqrt{45-66k+25k^2}}{2(2k-3)},~~~
w_3\equiv\frac{(k_2+2k_3)t^2}{2k-3},\nonumber
\end{eqnarray}
where $WM$ and $WW$ are Whittaker functions.
This tells us that for (\ref{fpowert}) alone one has Whittaker functions but adding $k=3/2$ 
times the shear the metric takes a power form.  To get Maple to compute (\ref{defalpha2}) the 
Whittaker expressions had to be done in steps,  a metric with Whittaker functions in it does 
not readily compute,  and for the $k=3/2$ case it was necessary to bring a constant factor 
into the power term.   $A_1$ in (\ref{fpoweras}) is in distributional form with standard 
deviation $\sqrt{3/((2(2k_2+k_3))}$,  taking all the $A$'s in this form but with different
standard deviations leads to a simple Ricci tensor which does not seem to be describable in
terms of simple matter fields and a Lanczos potential in integral form with integrand quartics
times times trignometric functions of quadratics. 

For possibilty \ref{qual}),
the geometric properties are:
transvected Bel-Robinson (\ref{bvdef})
\begin{eqnarray}
\label{genbv}
&&B_v=\frac{1}{6}
\left[(\dot{\beta_1}+\beta_1^2)^2
     +(\dot{\beta_2}+\beta_2^2)^2
     +(\dot{\beta_3}+\beta_3^2)^2
  +(\dot{\beta_1}+\beta_1^2)(-\beta_1\beta_2-\beta_1\beta_3+2\beta_2\beta_3)\right.\nonumber\\
&&+(\dot{\beta_2}+\beta_2^2)(-\beta_1\beta_2+2\beta_1\beta_3-\beta_2\beta_3)
  +(\dot{\beta_3}+\beta_3^2)(+2\beta_1\beta_2-\beta_1\beta_3-\beta_2\beta_3)\nonumber\\
&&-(\dot{\beta_1}+\beta_1^2)(\dot{\beta_2}+\beta_2^2)
  -(\dot{\beta_1}+\beta_1^2)(\dot{\beta_3}+\beta_3^2)
  -(\dot{\beta_2}+\beta_3^2)(\dot{\beta_3}+\beta_3^2)\nonumber\\
&&\left.-\beta_1^2\beta_2\beta_3-\beta_1\beta_2^2\beta_3-\beta_1\beta_2\beta_3^2
+\beta_1^2\beta_2^2+\beta_1^2\beta_3^2+\beta_2^2\beta_3^2\right],
\end{eqnarray}
transvected energy (\ref{hvdef})
\begin{eqnarray}
\label{genhv}
H_v&=&\frac{2}{9}(f_2^2+f_2f_3+f_3^2)
+\frac{2}{9}f_2(\beta_1-\beta_2)+\frac{2}{9}f_3(\beta_1-\beta_3)\\
&&+\frac{2}{27}\left(\beta_1^2+\beta_2^2+\beta_3^2-\beta_1\beta_2-\beta_1\beta_3-\beta_2\beta_3\right),
\nonumber
\end{eqnarray}
Weyl,  Ricci  and Lanczos scalars
\begin{eqnarray}
\label{genwr}
&&\Psi_0=\Psi_4=
-\frac{1}{4}\left(\beta_1(\beta_3-\beta_2)+\dot{\beta_2}+\beta_2^2-\dot{\beta_3}-\beta_3^2\right),\nonumber\\
&&\Psi_2=\frac{1}{12}\left(-2(\dot{\beta_1}+\beta_1^2)+\dot{\beta_2}+\beta_2^2+\dot{\beta_3}+\beta_3^2+\beta_1\beta_2+\beta_2\beta_3-2\beta_2\beta_3\right),\nonumber\\
&&\Phi_{00}=\Phi_{22}=-\frac{1}{4}\left(-\beta_1\beta_2-\beta_1\beta_3+\dot{\beta_2}+\beta_2^2+\dot{\beta_3}+\beta_3^2\right),~~~
\Phi_{11}=-\frac{1}{4}\left(\dot{\beta_1}+\beta_1^2-\beta_2\beta_3\right)\nonumber\\
&&\Phi_{02}=\Phi_{20}=-\frac{1}{4}\left(\beta_1(\beta_3-\beta_2)-\dot{\beta_2}-\beta_2^2+\dot{\beta_3}+\beta_3^2\right),\nonumber\\
&&\Phi_l=\frac{1}{12}\left(\dot{\beta_1}+\beta_1^2+\dot{\beta_2}+\beta_2^2+\dot{\beta_3}+\beta_3^2+\beta_1\beta_2+\beta_1\beta_3+\beta_2\beta_3\right),\nonumber\\
&&H_1=-H_6=-\frac{\sqrt{2}k}{36}\left(-2\beta_1+\beta_2+\beta_3\right)+\frac{\sqrt{2}}{12}(f_2+f_3),\nonumber\\
&&H_3=-H_4=\frac{\sqrt{2}k}{12}(\beta_2-\beta_3)+\frac{\sqrt{2}}{12}(f_3-f_2).
\end{eqnarray}
For $k=0$ the last two equations can be inverted to give the grand shear (\ref{newlc}) with
\begin{equation}
\label{invert}
f_1=-6\sqrt{2}H_1,~~~
f_2=3\sqrt{2}\left(H_1-H_3\right),~~~
f_3=3\sqrt{2}\left(H_1+H_3\right).
\end{equation}
\vspace{1cm}
\section{Cosmic energetics.}\label{ce}
What has been learnt about early cosmology seems to depend on ones point of view about 
gravitational energy.   The most straightforward approach is to say that gravitational energy
is related to the Weyl tensor and that there is Robertson-Walker geometry to way back early eras 
so there is not any.   There are grounds for think that the geometry is more complicated than
that (\cite{BKL},\cite{FB}) and that it becomes Levi-Cevita at early times,  this would 
introduce a non-vanishing Weyl tensor the changing nature of which has been conjectured
(\cite{penweyl},\cite{GH}).
If one assumes positive gravitational energy then gravitational energy just
adds to divergent quantities as one approaches the singularity.  If one assumes negative 
gravitational energy then there are intriguing possibilities such as it cancelling out matter
field energy and there being no energy at the singulatity and perhaps no overall energy at 
any time.  There could be an unbroken supersymmetry regime where the energies cancel.  
This suggest that the evolution of the 
universe could be thought of as exchange of energy between gravitation and matter fields.
Another way of looking at this is that observations tell us that the Universe is near the
critical value between open and closed,  the critical value could be thought of as having
no overall energy:  thus as the universe evolves energy transfers from Weyl tensor gravitational
energy to matter forms. 
Yet another way of looking at this is that if the universe started with a quantum flucuation
(\cite{tyron},\cite{mdr42}) then this was a departure from zero energy which set off everything.
Tranfer of energy suggests that entropy is non vanishing,  entropy tensors were first introduced 
in \cite{mdrcs}.
\newpage
\section{Conclusion.}\label{conc}
The equations for the Lanczos potential in Bianchi spacetime (\ref{grandequ}) have been set
up and solved in several cases although the general case remains intractable.
The equations (\ref{grandequ}) might have solutions with exotic properties.
A tensor which generalizes the shear called the grand shear is used as an intermediate step
in the calculation of the Lanczos potential, alternatively if a Lanczos potential and tetrad 
are known then the form of this tensor is immediate via (\ref{invert}).
The Lie derivative of the grand shear (\ref{liegen}) gives $1/t$ of the same object for 
Levi-Cevita (\ref{tdnewlc}) and vanishes for exponential spacetimes.
The grand shear might have application to other spacetimes.

Intuitively what has been learnt about the Lanczos potential is that the non-linearity of the
field equations has been 'transferred'.  The equations defining the Lanczos potential (\ref{ldef}) 
are linear,  but the solutions for the Lanzcos potential are often complicated - the 
non-linearity has been transfered to the potential.  This is hard to make precise because
the gauge can be changed (\ref{alggaugetrans}),  however if one takes fixed gauges then
(\ref{levicevitanonv}) is a cubic in $p$ over a quadratic in $p$ which is more complicated
than the quadratic in $p$ for the Riemann tensor (\ref{levicevitarie}).

Application to cosmic energetics raises more problems than it solves.
The energy tensors here do not seem to be directly related to quasi-local energy
(\cite{penquasi},\cite{WY}).
The good news is that rates of decay of all objects are what would be expected and wanted.
The form of (\ref{hess}) is what would be anticipated so there is a case for taking $H_1$
as the overall energy,  $\Psi_2$ is sometimes though of as a measure of mass/energy
but the two are not simply related as connection terms from covariant derivates arise 
when going from one to the other.
The transvected energy tensor (\ref{hvdef}) has awkward properties.
For Levi-Cevita spacetime it is independent of the $p$'s (\ref{lcvacen}),  which is no longer
the case (\ref{lcvacen2}) if the vacuum constraints (\ref{vacequal}) are not applied.
To see how the non-vacuum case is connected one transfers from $p$ constants to $\epsilon$
constants (\ref{changevar}):  taking the spatial $\epsilon_2$ to vanish just adds an $\epsilon$
term (\ref{e20}) with (\ref{hvdef}) still independent of $p$,  intuitively one has departed
from the vacuum by adding an energy $\epsilon_3$ and as this is monopolar it does not introduce
a $p$;  taking the temporal $\epsilon_3$ to vanish does introduce a $p$ (\ref{e30}),
intuitively this is a spatial departure from the vacuum introducing $p$.
For exponential space time (\ref{hvdef}) is constant (\ref{expyv}).
In general (\ref{genhv}) gives no indication of sign.
Thus for the transvected energy tensor (\ref{hvdef}) the sign and nature of any coupling
constant have not been fixed.

\end{document}